\documentclass[aps,prl,twocolumn,superscriptaddress,showpacs]{revtex4}
\usepackage{amsmath,amssymb}
\usepackage[dvips]{graphicx}
\usepackage{subfigure}
\usepackage{multirow}

\begin{document}
\preprint{arXiv:1310.1207}

\title{$\eta$ and $\eta^\prime$ mixing from Lattice QCD}

\newcommand\liv{Theoretical Physics Division, Department of Mathematical
  Sciences, The University of Liverpool, Liverpool L69 3BX, UK}
\newcommand\bn{Helmholtz Institut f{\"u}r Strahlen- und Kernphysik,
  Universit{\"a}t Bonn, Nussallee 14-16, 53115 Bonn, Germany}

\newcommand{\stat}{\mathrm{stat}}
\newcommand{\sys}{\mathrm{sys}}

\author{C.~Michael}\affiliation{\liv}
\author{K.~Ottnad}\affiliation{\bn}
\author{C.~Urbach}\affiliation{\bn}
\collaboration{ETM Collaboration}

\date{October 31, 2013}

\begin{abstract}
  We present a lattice QCD computation of $\eta$ and $\eta^\prime$
  masses and mixing angles, for the first time controlling continuum
  and quark mass extrapolations. The results for $M_\eta=
  551(8)_\stat(6)_\sys\ \mathrm{MeV}$ and $M_{\eta^\prime}=
  1006(54)_\stat(38)_\sys(+61)_\mathrm{ex}\ \mathrm{MeV}$ are in 
  excellent agreement with experiment. Our data show that the
  mixing in the quark flavour basis can be described by a single
  mixing angle of $\phi=46(1)_\stat(3)_\sys\,^\circ$ indicating that the
  $\eta^\prime$ is mainly a flavour singlet state.
\end{abstract}

\maketitle

\begin{figure}[t]
  \centering
  \includegraphics[width=.85\linewidth]%
  {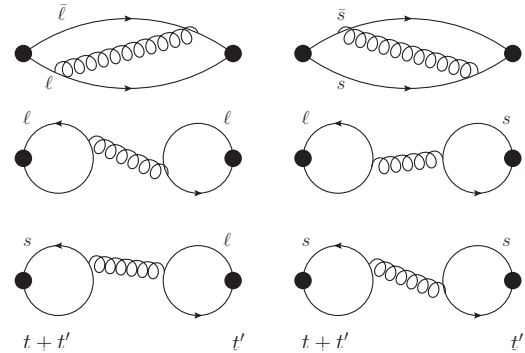}
  \caption{Light ($\ell$) and strange ($s$) connected (first row) and
    disconnected (second and third row) contributions from $t'$ to
    $t+t'$ for $\eta$ and
    $\eta'$ mesons. Curly lines symbolise the gluons.}
  \label{fig:wick}
\end{figure}

{\it Introduction.}---Quantum Chromodynamics (QCD) is established as
the theory of hadrons. Many of 
the properties of hadrons can be understood qualitatively from quark
models,  in which the gluons contribute only indirectly, being
responsible for the force  between quarks. A more direct consequence of
the  gluonic degrees of freedom in QCD is that quark loop contributions
(also known as disconnected diagrams or OZI-rule violating
contributions) are present -- see Figure~\ref{fig:wick} -- and they are
important in a description of the ninth pseudoscalar meson, the
$\eta'$ with mass $958\ \mathrm{MeV}$. In fact, if the disconnected
contributions were not present, the $\eta$ would have the mass of the
(neutral) pion. And the $\eta'$ would have a mass of
about $\sqrt{2M_\mathrm{K}^2-M_\pi^2}$, which is much lighter than the
physical $\eta'$ mass.

The large mass of the $\eta'$ meson is in contrast to the masses of 
the other eight light pseudoscalar mesons. Their small masses are
qualitatively explained by the spontaneous breaking of chiral symmetry
in QCD and the small quark mass values of up
($u$), down ($d$) and strange ($s$) quarks. The mass of the $\eta'$,
the ninth pseudoscalar meson, is thought to be caused by the
anomalous breaking of the $\mathrm{U}(1)_\mathrm{A}$ symmetry. This 
anomaly arises from the gluonic degrees of freedom in QCD and can be 
linked to the presence of topological excitations (of a pseudoscalar
nature) in the QCD vacuum
~\cite{Weinberg:1975ui,Belavin:1975fg,'tHooft:1976up}. 

This  scenario --based on arguments from effective field theory and
perturbation theory-- should be checked by a non-perturbative evaluation 
directly from QCD itself. The non-pertubative method which allows most
control over systematic errors is lattice QCD. Extrapolation to the
continuum limit requires computing  hadronic  properties for several
values of the lattice spacing ($a$).  Formalisms which have modifications
at finite  lattice spacing   of size $a^2$ allow this extrapolation to
be made  more reliably. Here we use the twisted mass lattice formalism
which  has this desirable property. It also has the attractive feature
of allowing  very efficient evaluation of disconnected contributions. To
study the  $\eta$ and $\eta'$, it is essential to include $u$, $d$ and
$s$ quarks. Here we go further and make use  of a twisted mass formalism
with the charm quark ($c$) also included.  Loop contributions from
even heavier quarks ($b$ and $t$) can safely be neglected.

Because of the essential contribution of disconnected diagrams, which 
are noisy when evaluated in lattice QCD, previous results for $\eta$ and
$\eta'$ using $u$, $d$ and $s$ quarks are quite
limited~\cite{Christ:2010dd,Kaneko:2009za,Dudek:2011tt,Gregory:2011sg,Dudek:2013yja}.
Previously, we have presented lattice results including also the $c$
quark in Ref.~\cite{Ottnad:2012fv} where we  were able to determine
the mass of the $\eta$ meson. Here we extend this study, using
powerful methods to separate ground and excited states, which will
enable us to determine $\eta$ and $\eta'$ mass and the corresponding
mixing angle, for the first time extrapolated to the continuum limit
and to physical values of up/down and strange quark masses.

This result is fundamental for our understanding of QCD and confirms the
role of the  ABJ anomaly in generating the observed large mass of the
$\eta'$ meson.

{\it Lattice computation.}---The results presented in this paper are based
on gauge configurations 
generated by the European Twisted Mass collaboration (ETMC) with
Iwasaki gauge action~\cite{Iwasaki:1985we} and Wilson twisted mass
fermions at maximal twist~\cite{Frezzotti:2000nk,Frezzotti:2003xj}
with up, down, strange and charm dynamical quark flavours. Up and down
quarks are mass degenerate and, therefore, mostly denoted with $\ell$
in the following. One
important advantage of this fermion action is that physical quantities
are $\mathcal{O}(a)$ improved~\cite{Frezzotti:2003ni}
(c.f. Refs.~\cite{Chiarappa:2006ae,Baron:2010bv}). The main drawback
of this action is that flavour and parity symmetries are 
broken at finite values of the lattice spacing, but restored in
the continuum limit.

The details of the three sets of ETMC ensembles (called $A$, $B$ and
$D$) were presented in Ref.~\cite{Baron:2010bv} and we adopt the notation 
therein for labeling the ensembles.
 A study of light meson masses and couplings (e.g. $f_{\pi}$), using 
chiral perturbation theory gave lattice spacings of $a_A=0.0863(4)\
\mathrm{fm}$, $a_B=0.0779(4)\ \mathrm{fm}$ and $a_D=0.0607(2)\
\mathrm{fm}$, respectively~\cite{Baron:2011sf}. This  can be summarised
conveniently by specifying the Sommer scale (which we measure accurately
on our lattices) as $r_0=0.45(2)$ fm, see
Refs.~\cite{Baron:2011sf,Ottnad:2012fv} for details. 
These ensembles cover a factor of two in $a^2$ so allow a reliable
determination  of the continuum limit ($a=0$). The physical volumes are,
with only a few exceptions, larger than $3\ \mathrm{fm}$ and $M_\pi
L\geq 3.5$. 

For each lattice spacing, the values of bare strange and charm quark
masses are kept fixed, while the bare average up/down quark mass value
is varied giving a lightest pseudoscalar meson (here called $M_{\pi}$)
in  the range $230-510$ MeV which allows reliable
extrapolation to the physical value. The value of the strange quark
mass is quite close to reproducing the physical kaon mass for the $B$
ensembles, and about 10\% too high for $A$ and $D$
ensembles~\cite{Baron:2011sf}. For two $A$-ensembles ($A80.24s$, $A100.24s$) 
we have results with a different strange quark mass and we use those to 
interpolate to the physical value of the kaon mass and to 
evaluate the small corrections to $M_\eta$. 
For more details on the ensemble parameters and number of configurations 
we refer to Ref.~\cite{Ottnad:2012fv}. 

For fixing average up/down and strange quark masses to their physical
values we use $M_{\pi^0}=135\ \mathrm{MeV}$ and $M_{\mathrm{K}^0}=498\
\mathrm{MeV}$. All statistical errors are determined using a blocked
bootstrap procedure with $1000$ samples to account for
autocorrelations. 

We compute the Euclidean correlation functions
\begin{equation}
  \label{eq:correlations}
  \mathcal{C}(t)_{qq'} =
  \langle\mathcal{O}_q(t'+t)\mathcal{O}_{q'}(t')\rangle\,,\quad
  q,q'\in{\ell,s,c}\,, 
\end{equation}
with operators $\mathcal{O}_\ell = (\bar ui\gamma_5 u + \bar d
i\gamma_5 d)/\sqrt{2}$, $\mathcal{O}_s = \bar s i\gamma_5 s$ and
$\mathcal{O}_c = \bar c i\gamma_5 c$. We enlarge our correlator matrix
$\mathcal{C}$ by including also fuzzed operators. Note that in twisted mass
lattice QCD there are several steps required to reach these
correlation functions, as explained in detail in
Ref.~\cite{Ottnad:2012fv}. We estimate the disconnected contributions
to the correlation functions Eq.~\ref{eq:correlations} using Gaussian
volume sources and the one-end trick for the connected
contributions~\cite{Boucaud:2008xu}. For the light disconnected
contributions a powerful noise reduction technique is
available~\cite{Jansen:2008wv,Ottnad:2012fv}. For the strange and
charm disconnected loops, we use the hopping parameter noise
reduction technique.

We solve the generalised eigenvalue problem
(GEVP)~\cite{Michael:1982gb,Luscher:1990ck,Blossier:2009kd}
\begin{equation}
  \label{eq:GEVP}
  \mathcal{C}(t)\eta^{(n)}(t,t_0) =
  \lambda^{(n)}(t,t_0)\mathcal{C}(t_0)\eta^{(n)}(t,t_0) 
\end{equation}
for eigenvalues $\lambda^{(n)}(t,t_0)$
and eigenvectors $\eta^{(n)}$. $n$ labels the states $\eta,
\eta^\prime,...$ contributing. Masses of these states can be determined
from the exponential fall-off of $\lambda^{(n)}(t,t_0)$ at large $t$.
The pseudoscalar matrix elements $A_{q,n}\equiv\langle
n|\mathcal{O}_q|0\rangle$ with $q \in \ell,s,c$ and $n\in
\eta,\eta',...$ can be extracted from the
eigenvectors~\cite{Blossier:2009kd}. It turns out that the charm quark
contributions to $\eta,\eta'$ are negligible and, thus, we drop
the $c$ quark in what follows.

\begin{figure}[t]
  \centering
  \subfigure[]{\includegraphics[width=.48\linewidth]
    {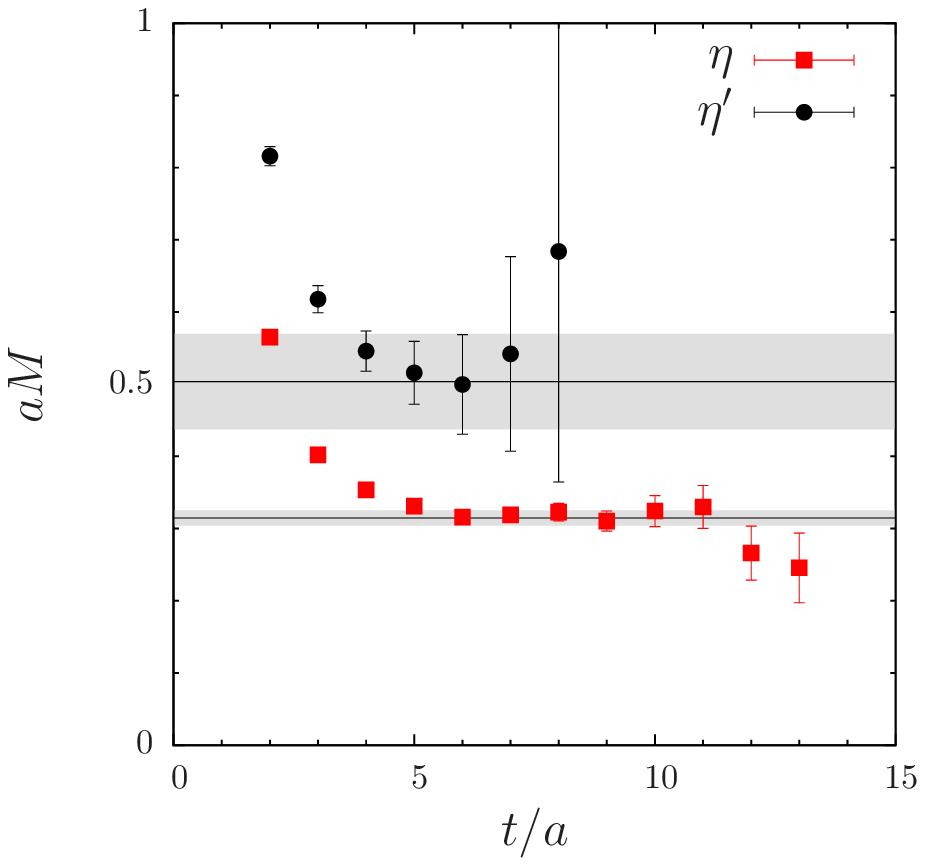}}\quad
  \subfigure[]{\includegraphics[width=.48\linewidth]
    {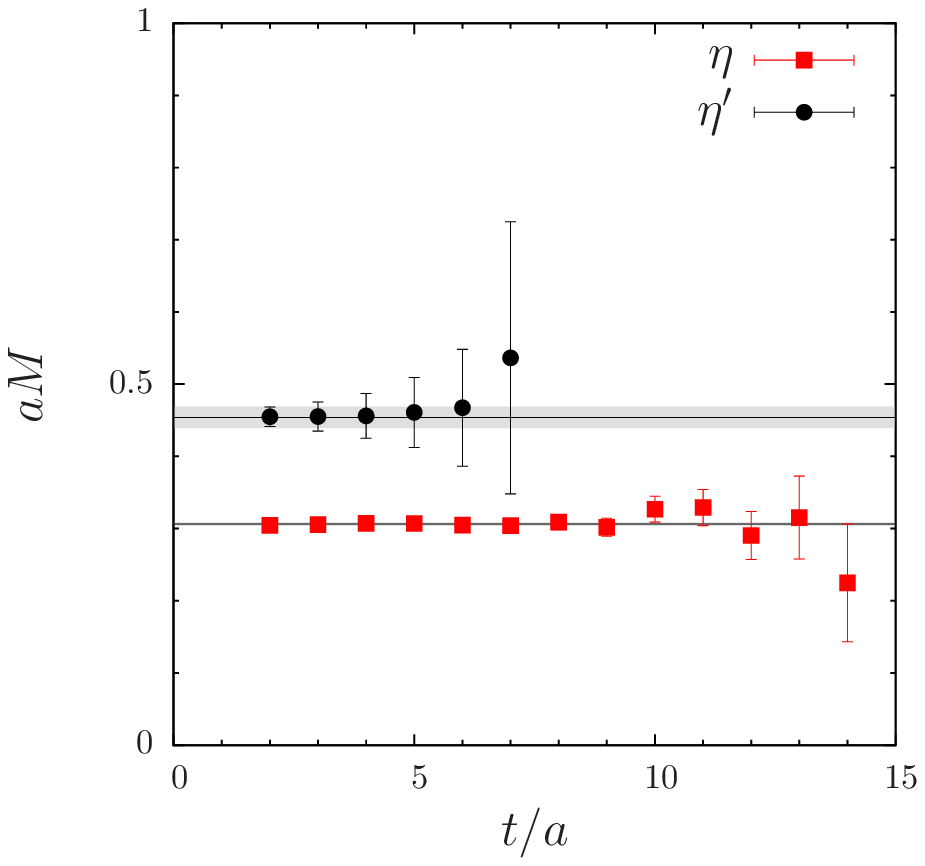}}\quad
  \caption{(a) Effective masses in lattice units determined from solving
    the GEVP for a $6\times6$ matrix with $t_0/a=1$ for ensemble
    A100. (b) the same as (a), but after removal of excited states in the
    connected contributions.}
  \label{fig:effmass}
\end{figure}

As an example for the masses determined from the GEVP we show in
Figure~\ref{fig:effmass}(a) the effective masses $aM^{(n)} =
-\log\{\lambda^{(n)}(t,t_0)/\lambda^{(n)}(t+1,t_0)\}$ as a function of
$t/a$ for ensemble $A100.24$. One observes a clear plateau
for the lowest state from $t/a=6$ on. For the first excited state, the
$\eta'$, a plateau is barely visible.

To improve the $\eta'$ (and $\eta$) mass determinations, we use a
method first proposed in Ref.~\cite{Neff:2001zr} and successfully
applied for the $\eta_2$ (the $\eta'$ in $N_f=2$ flavour QCD) in
Ref.~\cite{Jansen:2008wv}. The method is based on the following
assumption: disconnected contributions are only big for the $\eta$ and
$\eta'$ states, but negligible for higher excited states. The
assumption would be justified if topological charge fluctuations in
the vacuum, which give a large contribution to the $\eta'$ mass, 
mainly coupled to the $\eta, \eta'$ states, and not to other heavier
states.  Of course, the validity of
this assumption needs to be checked in the Monte-Carlo data.

The  connected contractions,  shown in Figure~\ref{fig:wick}, have a
constant signal to noise ratio in time, so we can reliably determine the
ground states in these two connected correlators and subtract the
excited state contributions. We then use these to build a correlation
matrix $\mathcal{C}^\mathrm{sub}$ from subtracted connected and original
disconnected contractions. 
 If disconnected contributions were relevant only for $\eta$ and
$\eta'$, one should find -- after diagonalising
$\mathcal{C}^\mathrm{sub}$ -- a plateau for both $\eta$ and $\eta'$ from
small values of $t/a$ on.

The effect of this procedure can be seen in Figure~\ref{fig:effmass}(b).
A plateau appears at much earlier $t/a$ values as compared to
Figure~\ref{fig:effmass}(a), while the plateau values agree very well
within errors. Therefore, we use this procedure -- which allows to
determine in particular the $\eta'$ mass value with much better accuracy
-- for the results presented here.

\begin{figure}[t]
  \centering
  \includegraphics[width=1.0\linewidth]{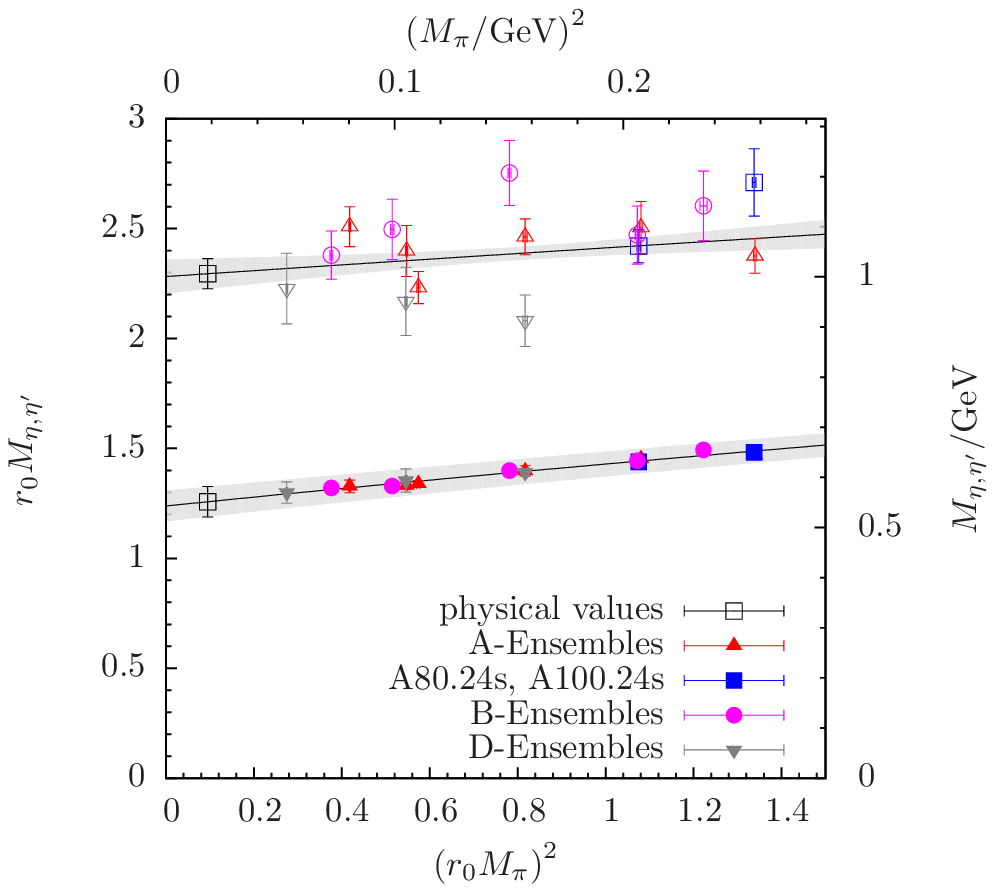}
  \caption{$\eta$ (filled) and $\eta'$ (open) masses versus
    $(r_0M_\pi)^2$ and chiral extrapolations with errorbands.}
  \label{fig:r0M}
\end{figure}

The $\eta$ and $\eta'$ mesons will be  flavour mixtures in general. For
the determination of the corresponding mixing angle, it is
convenient on the lattice to work in the quark flavour basis.
Our mixing determination builds on the pseudoscalar matrix elements $A_{q,n}$
which enter the fit to the correlators and which can be expressed with two
mixing angles $\phi_\ell,\phi_s$ and two constants $c_\ell, c_s$
 \begin{equation}
  \begin{pmatrix}
    A_{\ell, \eta} & A_{s, \eta}\\
    A_{\ell, \eta'} & A_{s, \eta'}\\
  \end{pmatrix}
  =
  \begin{pmatrix}
    c_\ell \cos\phi_\ell & -c_s \sin\phi_s \\
    c_\ell \sin\phi_\ell &  c_s \cos\phi_s \\
  \end{pmatrix}\,.
\end{equation}
 Given data for $A_{q,n}$, the mixing angles can be extracted:
 \begin{equation}
  \label{eq:angles}
  \tan\phi_\ell = \frac{A_{\ell,\eta'}}{A_{\ell,\eta}}\,
  ,\qquad\tan\phi_s = -\frac{A_{s,\eta}}{A_{s,\eta'}}\, . 
 \end{equation}
 Note that $c_\ell$, $c_s$ and renormalisation constants drop out in
Eqs.~\ref{eq:angles}. In chiral perturbation theory combined with large
$N_c$ arguments one can show that in the quark flavour basis $|\phi_s
-\phi_\ell|\ll 1$ should hold (see
Refs.~\cite{Schechter:1992iz,Kaiser:1998ds,Kaiser:2000gs,Feldmann:1998sh,Feldmann:1998vh} and references therein). If
this is the case, a single mixing angle  $\phi\approx\phi_s\approx\phi_\ell$
can be determined from
 \begin{equation}
  \label{eq:meanangle}
  \tan^2(\phi) = -\frac{A_{\ell\eta'}A_{s\eta}}{A_{\ell\eta}A_{s\eta'}}\,.
\end{equation}

{\it Results.}---To compare different lattice spacings, we plot  values
for $r_0M_\eta$ (corrected for any mismatch of the strange quark
mass~\cite{Ottnad:2012fv}) as filled symbols in Figure~\ref{fig:r0M},
with an error band dominated by the error of the strange quark mass
mismatch correction. The data from all three values of the 
lattice spacing fall onto a single line, indicating small lattice
artifacts and successful mismatch correction. The line represents a
linear fit of $(r_0M_\eta)^2$ in the 
squared pion mass to our data, resulting in $M_\eta = 551(11)_\stat\
\mathrm{MeV}$. Alternatively, we extrapolate $(M_\eta/M_\mathrm{K})^2$
or the GMO ratio $3M_\eta^2/(4M_\mathrm{K}^2-M_\pi^2)$ linearly in
the squared pion mass, leading to $M_\eta = 547(8)_\stat\ \mathrm{MeV}$ and 
$M_\eta=554(8)_\stat\ \mathrm{MeV}$, respectively. 
 We estimate the systematic uncertainties from fitting the different
lattice spacings separately and quote as a final result the weighted
average (accounting for correlations) over the three methods
\[
M_\eta = 551(8)_\stat(6)_\sys\ \mathrm{MeV}\,.
\]
 The data for $M_{\eta'}$ are plotted as open symbols in
Figure~\ref{fig:r0M} showing,  within the statistical errors, no visible
lattice spacing or strange quark mass dependence. Therefore,
we extrapolate the data of $(r_0M_{\eta'})^2$ for all
values of the lattice spacing, again linearly in the squared pion mass, to
the physical point and obtain
\[
M_{\eta'} = 1006(54)_\stat(38)_\sys(+61)_\mathrm{ex}\ \mathrm{MeV}\,,
\]
where again the systematic error comes from separate fits to the data
at the different lattice spacing values. In order to estimate the
potential systematic error stemming from the excited state removal
procedure, we use the difference of the extrapolations of data with and
without excited state removal.

It is worth noting that our estimates for $M_\eta$ and $M_{\eta'}$ are
in excellent agreement with the experimental values of $547.85(2)\
\mathrm{MeV}$ and $957.78(6)\ \mathrm{MeV}$~\cite{Beringer:1900zz},
respectively.

\begin{figure}[t]
  \centering
  \includegraphics[width=.85\linewidth]{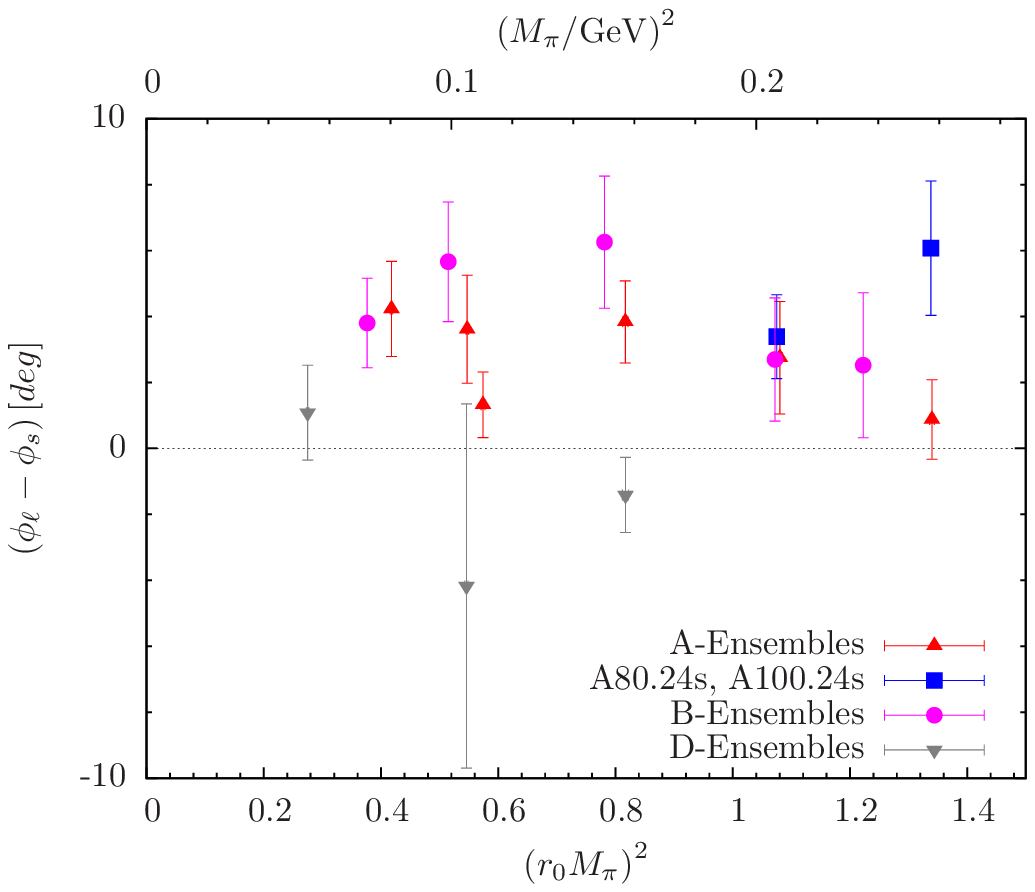}
  \caption{$\phi_\ell-\phi_s$ as a function of
    $(r_0M_\pi)^2$.}
  \label{fig:dphi}
\end{figure}

\begin{figure}[t]
  \centering
  \includegraphics[width=.85\linewidth]{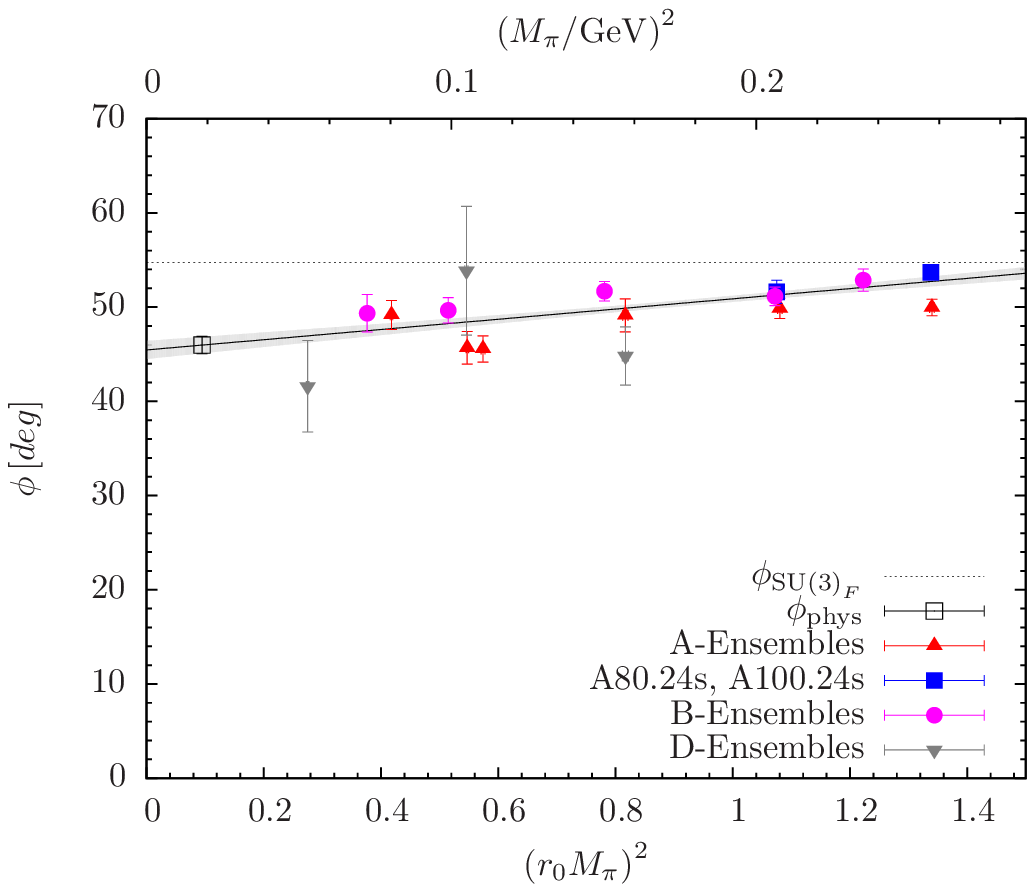}
  \caption{$\phi$ as a function of $(r_0M_\pi)^2$. Note that
    the data are consistent with the $\mathrm{SU}(3)$ flavour symmetry
    requirement
    $\phi_{\mathrm{SU}(3)_F}=\arctan{\sqrt{2}}\approx54.7^\circ$ for
    $m_\ell=m_s$.}
  \label{fig:phi}
\end{figure}

Using Eqs.~\ref{eq:angles}, we can evaluate $\phi_\ell$ and $\phi_s$ for
all ensembles. We first study the difference $\phi_\ell-\phi_s$ shown in
Figure~\ref{fig:dphi}: the difference is indeed rather small and
mostly compatible
with zero with  no significant dependence of the  strange quark mass.
 A linear extrapolation in
$(r_0M_\pi)^2$ of all the data yields
$3(1)_\stat(3)_\sys\,^\circ$, where the systematic error is estimated
from the maximal difference compared to extrapolating the data sets
for the three different lattice spacings separately.

In Figure~\ref{fig:phi} we show estimates for the single angle $\phi$
computed from Eq.~\ref{eq:meanangle}. Due to correlations of the data
used in the ratio, the statistical errors are smaller than
for $\phi_\ell$ and $\phi_s$ separately. The data appear to confirm
the smallness of $|\phi_\ell-\phi_s|$, as the overall picture is rather
consistent. Within the statistical accuracy, we cannot resolve (but
also not exclude) a residual lattice spacing and strange quark mass
dependence. Therefore, we perform a linear fit in the squared
pion mass to our data for $\phi$ and obtain
\[
\phi = 46(1)_\stat(3)_\sys\,^\circ\,.
\]
This value of $\phi$ indicates that the $\eta$ meson is dominated by the
flavour octet and the $\eta'$ mainly by the flavour singlet state.
It is in good agreement with other lattice
determinations~\cite{Christ:2010dd,Dudek:2011tt,Gregory:2011sg,Dudek:2013yja} and
slightly higher than the average phenomenological
estimate~\cite{Feldmann:1999uf}. We recall that we defined the angles
for the pseudoscalar densities. Due to the anomaly they are, therefore,
not directly related to the mixing angles defined via the axial vector
current.

We thank all members of ETMC for the most enjoyable collaboration. The
computer time for this project was made available to us by the John von
Neumann-Institute for Computing (NIC) on the JUDGE and Jugene
systems. In particular we thank U.-G.~Mei{\ss}ner for useful comments
and for granting us access on JUDGE. This project was funded by the
DFG as a project in the SFB/TR 16. K.O. and C.U. were supported by
the BCGS of Physics and Astronomie. The
open source software packages tmLQCD~\cite{Jansen:2009xp},
Lemon~\cite{Deuzeman:2011wz} and R~\cite{R:2005} have been used.

\bibliography{bibliography}

\end{document}